\documentclass[doublecol]{epl2}
\usepackage[english]{babel}
\usepackage{amssymb}
\usepackage{amsmath}
\usepackage{amssymb}
\usepackage{graphicx}
\usepackage{subfigure}
\usepackage{color}
\usepackage{mathrsfs}
\usepackage{soul}
\usepackage[T1]{fontenc} % Encoding package
\usepackage[dvipsnames]{xcolor}

\usepackage{hyperref}

\usepackage[utf8]{inputenc}

\usepackage{balance}
\usepackage{lastpage}

\title{Connections between kinks with different asymptotics}

\author{Vakhid A. Gani\inst{1,2}\footnote{E-mail: vagani@mephi.ru (corresponding author)} \and Aliakbar Moradi Marjaneh\inst{3}\footnote{E-mail: moradimarjaneh@gmail.com}}

\shortauthor{Vakhid A. Gani and Aliakbar Moradi Marjaneh}

\institute{                    
  \inst{1} National Research Nuclear University MEPhI (Moscow Engineering Physics Institute), Moscow 115409, Russia\\
  \inst{2} National Research Centre ``Kurchatov Institute'', Moscow 123182, Russia\\
  \inst{3} Department of Physics, Quchan Branch, Islamic Azad University, Quchan, Iran
}

\abstract{
In this letter, we show how to build bridges between field-theoretic models that have kink solutions with different asymptotic behavior. We study transformational properties of kinks in models with a real scalar field in two-dimensional space-time. We show how to obtain new models with solitons having different asymptotic behavior, using deformations with specific functions. In particular, starting from the well-known $\varphi^4$ model, we obtain kinks with super-exponential, super-super-exponential, power-law, and logarithmic asymptotics. We also comment on how asymptotic behavior of stability potential and zero mode can be obtained for deformed kinks.
}

\begin{document}

%\pacs{11.10.Lm, 11.27.+d, 05.45.Yv, 03.50.-z}

%11.10.Lm --- Nonlinear or nonlocal theories and models
%11.27.+d --- Extended classical solutions; cosmic strings, domain walls, texture,\\
%05.45.Yv --- Solitons,\\
%02.60.Cb --- Numerical simulation; solution of equations,\\
%02.30.Jr --- Partial differential equations,\\
%03.65.Pm --- Relativistic wave equations,\\
%03.50.-z --- Classical field theories,\\
%03.65.Ge --- Solutions of wave equations: bound states,\\

\maketitle

%\tableofcontents

\section{Introduction}
\label{sec:Introduction}

Kink solutions are found everywhere in physics \cite{Vilenkin.book.2000,Vachaspati.book.2006,Shnir.book.2018}. These solutions arise in (1+1)-dimensional field-theoretic models with a real scalar field and its self-interaction (potential term) with at least two degenerate minima. Kink solutions belong to the class of topological solitons \cite{Manton.book.2004}. There is an extensive literature on various properties of kinks, in particular, their pairwise (kink-antikink and kink-kink) interaction forces \cite{Christov.PRD.2019,Radomskiy.JPCS.2017,Christov.PRL.2019,Manton.JPA.2019,Campos.PLB.2021,dOrnellas.JPC.2020}, \cite[Sec.~10]{Khare.FP.2022}, the asymptotic properties of kink solutions \cite{Blinov.AoP.2022}, resonance phenomena in the kink-antikink scattering \cite{Campbell.PhysD.1986,Goodman.SIAM_JADS.2005,Anninos.PRD.1991,Gani.PRD.2014,Weigel.JPCS.2014,Bazeia.EPJC.2018.sinh,Yan.PLB.2020,Mohammadi.CNSNS.2021,Christov.CNSNS.2021,Alonso-Izquierdo.PRD.2021} and related features of the excitation spectra of individual kinks \cite{Gani.JHEP.2015,Gani.JPCS.2020.no-go} and kink+antikink systems as a whole \cite{Gani.JPCS.2020.no-go,Dorey.PRL.2011,Belendryasova.CNSNS.2019}.

An important direction in the study of properties of field-theoretic models and their kink solutions is the deformation procedure \cite{Blinov.AoP.2022,Bazeia.EPJC.2018.sinh,Bazeia.PRD.2002,Bazeia.PRD.2004,Bazeia.PRD.2006,Bazeia.PhysD.2008,Brito.AoP.2014,Moradi.CSF.2022,Khare.PLA.2022} --- a way to construct new models and kinks. The essence of this procedure is a special change of field variable that can be used to obtain new models with predetermined properties. Several interesting results have been obtained using the deformation procedure. Let us mention only some of them. In ref.~\cite{Bazeia.PRD.2004}, it was shown that it is possible to obtain a model with topological solutions (kinks) from a model with non-topological solutions (lumps). In ref.~\cite{Bazeia.PRD.2006}, various models with polynomial potentials in the form of Chebyshev polynomials were obtained as deformations of the $\varphi^4$ model \cite{Kevrekidis.book.2019,Belova.UFN.1997}. The sine-Gordon-like models were obtained from the $\varphi^4$ model in ref.~\cite{Bazeia.PhysD.2008}, and the properties of the new models and their topological solutions were investigated. The authors of ref.~\cite{Brito.AoP.2014} obtained multikink solutions from kinks of the $\varphi^4$ and the sine-Gordon models by tricky deformation. In ref.~\cite{Bazeia.EPJC.2018.sinh}, the process of kink-antikink scattering was studied in the model obtained from the $\varphi^4$ model using hyperbolic sine deformation. In comparison with analogous processes in the $\varphi^4$ model, fundamentally new phenomena have been discovered --- the formation of bound states of oscillons in the final state, as well as their (perhaps resonant) escape. Finally, in a recent paper \cite{Blinov.AoP.2022}, the changes in the asymptotic properties of kink solutions with respect to the deformation procedure were investigated. A class of deformations is considered under which power-law kink asymptotics remains power-law and exponential one remains exponential, but parameters of the asymptotics can change.

In this letter, we discuss deformations of field-theoretic models with one real scalar field in (1+1) dimensions. We develop a technology based on the deformation procedure for obtaining new models with kinks having new asymptotic properties. Namely, we are interested in deformations by singular functions that change the asymptotic behavior of kink solutions at spatial infinities. We show that such deformations allow us to obtain models with kinks having asymptotics faster than exponential ones --- for example, super-exponential and super-super-exponential, as well as slower than exponential ones --- for example, power-law and logarithmic.

The physical motivation for our study comes from numerous applications of kinks with different asymptotic behavior, see, {\it e.g.}, refs.~\cite{Khare.FP.2022,Khare.PS.2019,Khare.JPA.2020}. In the last decade, the relevance of studying the interaction forces between kink configurations has increased. These forces critically depend on the behavior of the field as a function of the spatial coordinate far from the location of the kink, {\it i.e.} on its asymptotics. For example, in refs.~\cite{Khare.JPA.2019,Blinov.CSF.2022} many new kink solutions with power-law asymptotic behavior were found, while in refs.~\cite{Christov.PRD.2019,Radomskiy.JPCS.2017,Christov.PRL.2019,Manton.JPA.2019,Campos.PLB.2021,dOrnellas.JPC.2020} the interactions of kinks with power-law tails were investigated, see also a comprehensive review \cite{Khare.FP.2022}.

We illustrate our method by deformations of the well-known $\varphi^4$ model. The $\varphi^4$ model is remarkable because the history of intensive studying its kink solutions dates back more than half a century; see, {\it e.g.}, reviews \cite{Kevrekidis.book.2019,Belova.UFN.1997}. In collisions of kinks of this model, resonant phenomena were discovered --- the so-called escape windows caused by the resonant energy exchange between the translational and vibrational modes of the kink. Pioneer works on this topic date back to the 70s of the last century, see, {\it e.g.}, refs.~\cite{Kudryavtsev.JETPLett.1975,Sugiyama.PTP.1979}. Nevertheless, kinks of the $\varphi^4$ model and their interactions are still being actively studied \cite{Mohammadi.CNSNS.2021,Adam.PRE.2020,Manton.PRD.2021}.

Our paper is organized as follows. In section ``Kink solutions'', we first briefly recall the basic information about kink solutions. Then, in section ``Deformations'' we recall the basics about deformations. Section ``Change of the asymptotic behavior'' introduces a deformation function with the properties we need. In this section we find how the asymptotic behavior of the potential and of the kink solution change under such deformations. In section ``Example of the $\varphi^4$ model'', an example of a specific deforming function applied to the $\varphi^4$ model is analyzed. In section ``A remark on the stability potential'', we explain how asymptotical behavior of the kink stability potential and kink's zero mode can be obtained. Finally, we conclude in section ``Conclusion''.

\section{Kink solutions}
\label{sec:Kink_solutions}

First, let us briefly outline the context. Consider a field-theoretic model with a real scalar field living in $(1+1)$-dimensional space-time, defined by the Lagrangian
\begin{equation}\label{eq:Lagrangian}
    \mathcal{L} = \frac{1}{2}\left(\frac{\partial \varphi}{\partial t}\right)^2-\frac{1}{2}\left(\frac{\partial \varphi}{\partial x}\right)^2-V(\varphi),
\end{equation}
where $V(\varphi)$ is the self-interaction of the field ({\it i.e.}, potential of the model) --- real-valued non-negative function, having at least two degenerate minima.

The equation of motion for the Lagrangian \eqref{eq:Lagrangian} is the nonlinear Klein--Gordon equation:
\begin{equation}\label{eq:eqmo}
    \frac{\partial^2\varphi}{\partial t^2} - \frac{\partial^2\varphi}{\partial x^2} + \frac{dV}{d\varphi} = 0.
\end{equation}
A static kink is a field configuration $\varphi_{\rm K}^{}(x)$, which monotonously interpolates between neighboring minima $\varphi_1^{}$ and $\varphi_2^{}$ of the potential $V(\varphi)$ ({\it i.e.}, $\lim\limits_{x\to-\infty}\varphi_{\rm K}^{}(x)=\varphi_1^{}$, $\lim\limits_{x\to+\infty}\varphi_{\rm K}^{}(x)=\varphi_2^{}$) and satisfies the Bogomolny--Prasad--Sommerfield equation \cite{BPS1,BPS2}:
\begin{equation}\label{eq:BPS}
    \frac{d\varphi}{dx} = \sqrt{2V(\varphi)},
\end{equation}
for more details see, {\it e.g.}, \cite[Sec.~II]{Gani.PRD.2014} or \cite[Sec.~2]{Gani.JHEP.2015}

For example, in the case of the potential $V(\varphi)=\frac{1}{2}\left(1-\varphi^2\right)^2$ we have the so-called $\varphi^4$ model \cite{Kevrekidis.book.2019} with the kink $\varphi_{\rm K}^{}(x) = \tanh x$. We can say with confidence that this model is of particular importance for theoretical and mathematical physics. In addition to its numerous applications, ranging from the old phenomenological Ginzburg--Landau theory \cite{Landau.ZhETF.1937,Ginzburg.ZhETF.1950,Tinkham.book.1996} to modern studies of phenomena in graphene \cite{Yamaletdinov.PRB.2017}, the model is also remarkable in that historically the studies of the dynamics of kink-antikink collisions began with $\varphi^4$ kinks \cite{Kudryavtsev.JETPLett.1975}, see also reviews \cite{Kevrekidis.book.2019,Belova.UFN.1997}.

Below we will discuss the asymptotic behavior of kinks. By asymptotics we will understand how the function $\varphi_{\rm K}^{}(x)$ approaches the vacuum value of the field $\varphi$ at $x\to-\infty$ or $x\to+\infty$. For example, it is obvious that the left and right asymptotics of the $\varphi^4$ kink are exponential.

\section{Deformations}
\label{sec:Deformations}

Deformation procedure \cite{Bazeia.PRD.2002} is a change of the field variable, which enables, starting from a model with known kink solution, to obtain a new model along with its kink(s). The substantial advantage of the deformation procedure is that one can obtain new models with predefined properties.

Let us briefly recall the essence of the deformation procedure. We start with a field-theoretic model with the potential $V^{(0)}(\varphi)$ and the kink $\varphi_{\rm K}^{(0)}(x)$. To implement the deformation, a deformation function $f(\varphi)$ is specified, which determines the change of variable and has the required properties. The potential and kink of the deformed model have the form
\begin{equation}\label{eq:f-deformation}
    V^{(1)}(\varphi) = \frac{V^{(0)}[f(\varphi)]}{[f^\prime(\varphi)]^2}
\end{equation}
and
\begin{equation}\label{eq:new_model_kink1}
    \varphi_{\rm K}^{(1)}(x) = f^{-1}[\varphi_{\rm K}^{(0)}(x)].
\end{equation}
If the kink $\varphi_{\rm K}^{(0)}(x)$ connects the vacua $\varphi_1^{}$ and $\varphi_2^{}$ of the original model, the kink $\varphi_{\rm K}^{(1)}(x)$ connects the vacua $f^{-1}(\varphi_1^{})$ and $f^{-1}(\varphi_2^{})$ of the deformed model.

In ref.~\cite{Blinov.AoP.2022} it was shown that the deformation procedure can also be applied in the case of implicit kinks, {\it i.e.}, when the kink of the original model is known as $x=x_{\rm K}^{(0)}(\varphi)$. The deformed kink then looks like
\begin{equation}\label{eq:first_deform_implicit_kink}
x_{\rm K}^{(1)}(\varphi)=x_{\rm K}^{(0)}(f(\varphi)).
\end{equation}
For details on the deformation procedure see, {\it e.g.}, refs.~\cite{Blinov.AoP.2022,Bazeia.EPJC.2018.sinh,Bazeia.PRD.2002,Bazeia.PRD.2004}. In addition, in ref.~\cite{Blinov.AoP.2022} the question of how the kink asymptotics changes when the model is deformed was studied. It was shown that, depending on the properties of the deformation function, the kink asymptotics can change to a greater or lesser extent. At the same time, ref.~\cite{Blinov.AoP.2022} considers only deformation functions for which the exponential kink asymptotics remains exponential, while the power-law asymptotics remains power-law. In this letter, we will discuss deformations which change the kink asymptotics drastically: for example, the exponential asymptotics can change to a power-law one  (which is much slower than exponential), or, conversely, into a super-exponential one (which is much faster than exponential).

\section{Change of the asymptotic behavior}
\label{sec:Change_of_the_asymptotic_behavior}

Assume that the potential $V^{(0)}(\varphi)$ of the original model is a non-negative function having two or more degenerate minima where it vanishes. Assume that at $\varphi\to\varphi_0^{}-0$ the potential can be approximated by
\begin{equation}\label{eq:polynomial_potential_0_approx}
    V^{(0)}(\varphi) \approx \frac{1}{2}\left(\varphi_0^{}-\varphi\right)^{2n} v(\varphi_0^{}),
\end{equation}
where $v(\varphi_0^{})>0$. For example, $V^{(0)}(\varphi)$ can be a polynomial, which can be factorized as
\begin{equation}\label{eq:polynomial_potential_0}
    V^{(0)}(\varphi) = \frac{1}{2}\left(\varphi_0^{}-\varphi\right)^{2n} v(\varphi),
\end{equation}
where $n$ is a positive integer, and $v(\varphi)$ is a non-negative polynomial of even degree, $v(\varphi_0^{})>0$. Note that the below considerations can also be applied to the potential of the form \eqref{eq:polynomial_potential_0_approx} with $v(\varphi_0^{})$ being replaced by $v(\varphi_0^{}-\varphi)$ having no (finite) limit at $\varphi\to\varphi_0^{}-0$.

If $n=1$ then the corresponding asymptotics of the kink is exponential:
\begin{equation}\label{eq:exponential_asymptotics_0}
    \varphi_{\rm K}^{(0)}(x) \approx \varphi_0^{} - \exp\left[-\sqrt{v(\varphi_0^{})}\:x\right] \quad \mbox{at} \quad x\to+\infty.
\end{equation}
For $n\ge 2$ the kink asymptotics is power-law:
\begin{equation}\label{eq:kink_asymptotics_0}
    \varphi_{\rm K}^{(0)}(x) \approx \varphi_0^{} - \frac{A_n^{(0)}}{x^{1/(n-1)}} \quad \mbox{at} \quad x\to+\infty,
\end{equation}
where
\begin{equation}\label{eq:A_0}
     A_n^{(0)} = \left[\left(n-1\right)\sqrt{v(\varphi_0^{})}\right]^{1/(1-n)}.
\end{equation}
(For more details, see ref.~\cite{Blinov.AoP.2022}.)

Now suppose that the original model is deformed using a deformation function that has the following properties:
\begin{enumerate}
    \item[(i)] the function $f(\varphi)$ strictly monotonically increases in the interval between two vacua of the model, of which $\varphi_0^{}$ is located on the right;
    \item[(ii)] $f(\varphi)\approx\varphi_0^{} - C \exp\left(-\displaystyle\frac{1}{\varphi_0^{}-\varphi}\right)$ and $f^\prime(\varphi)\approx\displaystyle\frac{C}{\left(\varphi_0^{}-\varphi\right)^2}\exp\left(-\displaystyle\frac{1}{\varphi_0^{}-\varphi}\right)$ at $\varphi\to\varphi_0^{}-0$.
\end{enumerate}
Note that $\varphi_0^{}$ is a fixed point of the deformation function in the sense that $\lim\limits_{\varphi\to\varphi_0^{}-0}f(\varphi)=\varphi_0^{}$.

The deformed potential \eqref{eq:f-deformation} then behaves at $\varphi\to\varphi_0^{}-0$ as
\begin{equation}
    V^{(1)}(\varphi) \approx \frac{1}{2}C^{2n-2}v\left(\varphi_0^{}\right)\left(\varphi_0^{}-\varphi\right)^4 \exp\left(-\frac{2n-2}{\varphi_0^{}-\varphi}\right).
%    \quad \mbox{at} \quad \varphi\to\varphi_0^{}-0.
\end{equation}
For $n\ge 2$ the saymptotics of kink $\varphi_{\rm K}^{(1)}(x)$ at $x\to+\infty$ in such potential is
\begin{equation}
    \varphi_{\rm K}^{(1)}(x) \approx \varphi_0^{} - \frac{n-1}{\ln\left[\left(n-1\right)\sqrt{v(\varphi_0^{})}\:x\right]},
%    \quad \mbox{at} \quad x\to+\infty.
\end{equation}
while for $n=1$ the asymptotics of $\varphi_{\rm K}^{(1)}(x)$ is power-law. At the same time, the potential $V^{(0)}(\varphi)$ can also be deformed by the inverse function $f^{-1}(\varphi)$ with the resulting potential $V^{(-1)}(\varphi)$ at $\varphi\to\varphi_0^{}-0$:
\begin{equation}
    V^{(-1)}(\varphi) \approx \frac{1}{2} v\left(\varphi_0^{}\right)\left(\varphi_0^{}-\varphi\right)^2 \left[ \ln\left(\frac{C}{\varphi_0^{}-\varphi}\right)\right]^{4-2n}.
%    \quad \mbox{at} \quad \varphi\to\varphi_0^{}-0.
\end{equation}
For example, for $n\ge 2$ the kink $\varphi_{\rm K}^{(-1)}(x)$ in such potential has the following asymptotics at $x\to+\infty$:
\begin{equation}\nonumber
    \varphi_{\rm K}^{(-1)}(x) \approx
\end{equation}
\begin{equation}
%    \varphi_{\rm K}^{(-1)}(x) \approx
    \varphi_0^{} - C\exp\left\{-\left[\left(n-1\right)\sqrt{v(\varphi_0^{})}\right]^{1/(n-1)} x^{1/(n-1)}\right\}.
%    \quad \mbox{at} \quad x\to+\infty.
\end{equation}
Of course, the deformation can be continued, on the one hand, by repeated application of the function $f$, resulting in $V^{(2)}(\varphi)$ and so on, and on the other hand, by repeated application of the function $f^{-1}$, resulting in $V^{(-2)}(\varphi)$ and so on.

As an example, below we apply the function $f(\varphi) = \tanh\left(\displaystyle\frac{\varphi}{1-\varphi^2}\right)$ and its inverse to the well-known $\varphi^4$ model.

\section{Example of the \texorpdfstring{$\varphi^4$}{pdfbookmark} model}
\label{sec:Example_of_the_phi4_model}

As it has already been mentioned above, the potential of the $\varphi^4$ model and its kink, respectively, are
\begin{equation}\label{eq:phi4_potential_and_kink}
    V^{(0)}(\varphi)=\frac{1}{2}\left(1-\varphi^2\right)^2 \quad  \mbox{and} \quad \varphi_{\rm K}^{(0)}(x) = \tanh x.
\end{equation}
%
%and
%
%\begin{equation}\label{eq:phi4kink}
%    \varphi_{\rm K}^{(0)}(x) = \tanh x.
%\end{equation}
%
%The $\varphi^4$ kink is symmetric, its mass is 4/3. The stability potential of this kink is the modified Poeschl--Teller potential \cite{PT} with two discrete levels: zero mode $\omega_0^{}=0$ and vibrational mode $\omega_1^{}=\sqrt{3}$.
The left and right asymptotics of the $\varphi^4$ kink are exponential:
%
%\begin{equation}\label{eq:phi4_asymptotics}
%    \varphi_{\rm K}^{(0)}(x) \approx
%    \begin{cases}
%        -1 + e^{2x} \quad \mbox{at} \quad x\to-\infty,\\
%        \thinspace 1 - e^{-2x} \quad \mbox{at} \quad x\to+\infty.
%    \end{cases}
%\end{equation}
%
\begin{equation}\label{eq:phi4_asymptotics}
    \varphi_{\rm K}^{(0)}(x) \approx
    \pm 1 \mp 2 e^{\mp 2x} \quad \mbox{at} \quad x\to \pm\infty.
\end{equation}

Let us show how by means of deformation of the $\varphi^4$ model we can obtain new models with kinks with a completely different asymptotic behavior. As deformation function we take
\begin{equation}\label{eq:f}
    f(\varphi) = \tanh\left(\frac{\varphi}{1-\varphi^2}\right),
\end{equation}
and its inverse
\begin{equation}\label{eq:inverse_f}
    f^{-1}(\varphi) = \frac{ \sqrt{1+\ln^2\frac{1+\varphi}{1-\varphi}}-1}{\ln \frac{1+\varphi }{1-\varphi }}.
\end{equation}
%
%see Fig.~\ref{fig:ffinv}.
%
%\begin{figure*}[t!]
%\begin{center}
%  \centering
%  \subfigure[]{\includegraphics[width=0.45
% \textwidth]{f.pdf}\label{fig:f}}
%% \hspace{1mm}
% \hfill
%  \subfigure[]{\includegraphics[width=0.45
% \textwidth]{inverse_f.pdf}\label{fig:inverse_f}}
%%  \hspace{1mm}
%  \caption{Deformation function \eqref{eq:f} and its inverse \eqref{eq:inverse_f}.}
%  \label{fig:ffinv}
%\end{center}
%\end{figure*}

Two consecutive applications of the function $f(\varphi)$ \eqref{eq:f} to the $\varphi^4$ model \eqref{eq:phi4_potential_and_kink} give
\begin{equation}\label{eq:f-deformed_potential}
    V^{(1)}(\varphi) = \frac{1}{2} \cdot \frac{\left(1-\varphi^2\right)^4}{\left(1+\varphi^2\right)^2}
%    \quad  \mbox{and} \quad \varphi_{\rm K}^{(1)}(x) = \frac{\sqrt{1+4x^2}-1}{2x},
\end{equation}
and
\begin{equation}\label{eq:f-deformed_kink}
%    V^{(1)}(\varphi) = \frac{1}{2} \cdot \frac{\left(1-\varphi^2\right)^4}{\left(1+\varphi^2\right)^2} \quad  \mbox{and} \quad
    \varphi_{\rm K}^{(1)}(x) = \frac{\sqrt{1+4x^2}-1}{2x},
\end{equation}
\begin{equation}\label{eq:ff-deformed_potential}
    V^{(2)}(\varphi) = \frac{1}{2} \cdot \frac{\left(1-\varphi^2\right)^4}{\left(1+\varphi^2\right)^2} \cdot \frac{1}{\cosh^2\frac{2\varphi}{1-\varphi^2}}
%    \quad  \mbox{and} \quad \varphi_{\rm K}^{(2)}(x) = \frac{\sqrt{1+\ln^2\left(2x+\sqrt{1+4x^2}\right)}-1}{\ln\left(2x+\sqrt{1+4x^2}\right)}.
\end{equation}
and
\begin{equation}\label{eq:ff-deformed_kink}
%    V^{(2)}(\varphi) = \frac{1}{2} \cdot \frac{\left(1-\varphi^2\right)^4}{\left(1+\varphi^2\right)^2} \cdot \frac{1}{\cosh^2\frac{2\varphi}{1-\varphi^2}} \quad  \mbox{and} \quad
    \varphi_{\rm K}^{(2)}(x) = \frac{\sqrt{1+\ln^2\left(2x+\sqrt{1+4x^2}\right)}-1}{\ln\left(2x+\sqrt{1+4x^2}\right)}.
\end{equation}
On the other hand, consecutively applying the function $f^{-1}(\varphi)$ \eqref{eq:inverse_f} to \eqref{eq:phi4_potential_and_kink}, we obtain
\begin{equation}\label{eq:f-1-deformed_potential}
    V^{(-1)}(\varphi) = \frac{1}{2} \left(1-\varphi^2\right)^2 \left(1+\ln^2\frac{1+\varphi}{1-\varphi}\right)
%    \quad  \mbox{and} \quad \varphi_{\rm K}^{(-1)}(x) = \tanh\left[\frac{1}{2}\sinh\left(2x\right)\right],
\end{equation}
and
\begin{equation}\label{eq:f-1-deformed_kink}
%    V^{(-1)}(\varphi) = \frac{1}{2} \left(1-\varphi^2\right)^2 \left(1+\ln^2\frac{1+\varphi}{1-\varphi}\right) \quad  \mbox{and} \quad
    \varphi_{\rm K}^{(-1)}(x) = \tanh\left[\frac{1}{2}\sinh\left(2x\right)\right],
\end{equation}
\begin{equation}\nonumber
    V^{(-2)}(\varphi) = \frac{1}{2} \left(1-\varphi^2\right)^2 \left(1+\ln^2\frac{1+\varphi}{1-\varphi}\right)\times
\end{equation}
\begin{equation}\label{eq:f-1f-1-deformed_potential}
    \left( 1 + \ln^2 \frac{\ln\frac{1+\varphi}{1-\varphi} + \sqrt{1+\ln^2\frac{1+\varphi}{1-\varphi}}-1}{\ln\frac{1+\varphi}{1-\varphi} - \sqrt{1+\ln^2\frac{1+\varphi}{1-\varphi}}+1}\right)
\end{equation}
and
\begin{equation}\label{eq:f-1f-1-deformed_kink}
    \varphi_{\rm K}^{(-2)}(x) = \tanh\left[\frac{1}{2}\sinh\left(\sinh(2x)\right)\right].
\end{equation}

\begin{figure*}[t!]
\begin{center}
  \centering
  \subfigure[]{\includegraphics[width=0.45
 \textwidth]{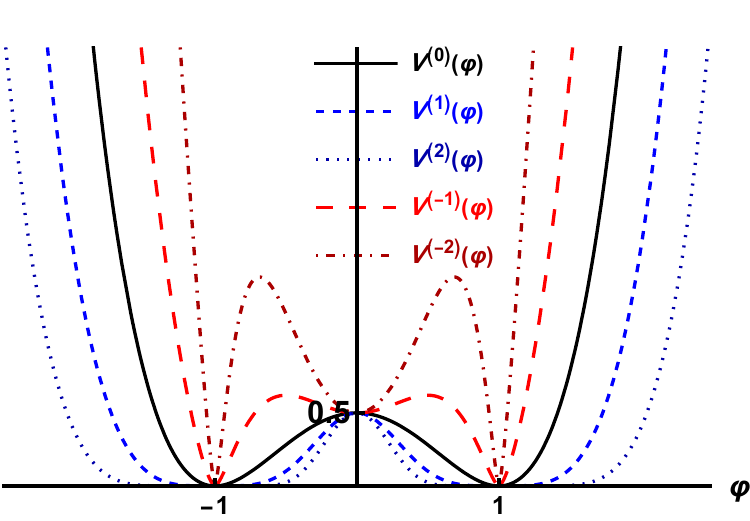}\label{fig:potentials}}
%  \hspace{1mm}
  \hfill
  \subfigure[]{\includegraphics[width=0.45
 \textwidth]{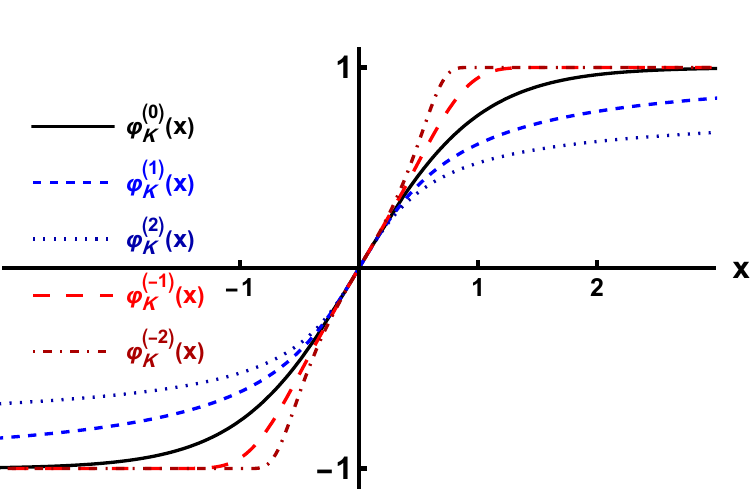}\label{fig:kinks}}
%  \hspace{1mm}
  \caption{Potentials and kinks of the models obtained by deformation of the $\varphi^4$ model.}
    \label{fig:PotentialsKinks}
\end{center}
\end{figure*}

The above potentials and kinks are shown in fig.~\ref{fig:PotentialsKinks}.
Note that the potentials $V^{(-1)}(\varphi)$ and $V^{(-2)}(\varphi)$, eqs.~\eqref{eq:f-1-deformed_potential} and \eqref{eq:f-1f-1-deformed_potential}, are defined on the interval $-1<\varphi<1$. Such a narrowing of the domain of definition of the potential is a consequence of the deformation by $f^{-1}$. Since the kink solution is strictly monotonic and $-1<\varphi_{\rm K}^{(-1)}(x)<1$ for all $x$, no problem arises. However, when modeling a real physical situation, one should apparently extend the domain of definition of the potential $V^{(-1)}(\varphi)$ to $|x|\ge 1$, {\it e.g.}, by using $\displaystyle\left|\frac{1+\varphi}{1-\varphi}\right|$ instead of $\displaystyle\frac{1+\varphi}{1-\varphi}$. This is necessary because, as a result of fluctuations, the field can go beyond the interval $-1<\varphi<1$.

Thus, by specifying the initial field-theoretic model and the deforming function, we obtain a tower of models and their kinks having different asymptotics. Let us now write out and discuss the asymptotic behavior of all the above kinks (all kinks are symmetric, hence we give the right asymptotics only). Continuing to use the notation accepted here for deformed kinks, we obtain
\begin{itemize}
    \item the power-law asymptotics of $\varphi_{\rm K}^{(1)}(x)$:
    \begin{equation}\label{eq:f-deformed_kink_asymptotics}
        \varphi_{\rm K}^{(1)}(x) \approx 1 - \frac{1}{2x} \quad \mbox{at} \quad x\to +\infty;
    \end{equation}
    \item the logarithmic asymptotics of $\varphi_{\rm K}^{(2)}(x)$:
    \begin{equation}\label{eq:ff-deformed_kink_asymptotics}
        \varphi_{\rm K}^{(2)}(x) \approx 1 - \frac{1}{\ln\left(4x\right)} \quad \mbox{at} \quad x\to +\infty.
    \end{equation}
    \item the super-exponential asymptotics of $\varphi_{\rm K}^{(-1)}(x)$:
    \begin{equation}\label{eq:f-1-deformed_kink_asymptotics}
        \varphi_{\rm K}^{(-1)}(x) \approx 1 - 2 e^{-\frac{1}{2}e^{2x}} \quad \mbox{at} \quad x\to +\infty;
    \end{equation}
    \item the super-super-exponential asymptotics of $\varphi_{\rm K}^{(-2)}(x)$:
    \begin{equation}\label{eq:f-1f-1-deformed_kink_asymptotics}
        \varphi_{\rm K}^{(-2)}(x) \approx 1 - 2 e^{-\frac{1}{2}e^{e^{x}}} \quad \mbox{at} \quad x\to +\infty;
    \end{equation}
\end{itemize}

This example demonstrates that successive application of the deformation function $f$ slows down the kink asymptotics, {\it i.e.} delocalizes the kink. At the same time, successive application of the deformation function $f^{-1}$, on the contrary, accelerates the kink asymptotics, the asymptotic approach of the field to the vacuum value becomes faster with distance from the kink center. We can assume that in the limit of an (infinitely) large number of successive deformations by the function $f^{-1}$ we obtain a compact kink. However, this requires further study, which is beyond the scope of this paper.

Note that if we apply the same deformation function \eqref{eq:f} to the $\varphi^6$ model with potential $V(\varphi)=\frac{1}{2}\varphi^2(1-\varphi^2)^2$, then we obtain the results similar to the case of deformations of the $\varphi^4$ model, except that for the asymmetric $\varphi^6$ kink, only the asymptotic behavior near the vacuum 1 changes significantly, while the asymptotic behavior near the vacuum 0 does not change so significantly.

\section{A remark on the stability potential}
\label{sec:A_remark_on_the_stability_potential}

In many processes involving kinks, the presence or absence of so-called vibrational modes --- internal vibrational degrees of freedom that can be excited in certain situations --- is important. For example, the accumulation of energy by the vibrational modes of the kink and antikink explains the presence of escape windows (sometimes called bounce windows) and quasi-resonances (sometimes called false windows) in kink-antikink collisions \cite{Campbell.PhysD.1986,Goodman.SIAM_JADS.2005,Anninos.PRD.1991,Gani.PRD.2014,Weigel.JPCS.2014,Bazeia.EPJC.2018.sinh,Yan.PLB.2020,Mohammadi.CNSNS.2021,Christov.CNSNS.2021,Alonso-Izquierdo.PRD.2021,Gani.JHEP.2015}. Moreover, it was found that even in the absence of vibrational modes in isolated kinks and antikinks, collective vibrational modes can appear when the kink and antikink approach each other \cite{Dorey.PRL.2011,Belendryasova.CNSNS.2019}. The presence or absence of vibrational modes in the excitation spectrum of a kink critically depends on the asymptotic behavior of the kink. For example, a kink with at least one power-law asymptotic cannot have a vibrational mode \cite{Blinov.AoP.2022,Gani.JPCS.2020.no-go}.

The spectrum of small excitations of the kink in the linear approximation can be obtained as follows. We add a small perturbation $\delta\varphi(x,t)$ to the static kink $\varphi_{\rm K}^{}(x)$, and substitute the configuration $\varphi(x,t) = \varphi_{\rm K}^{}(x) + \delta\varphi(x,t)$ into the equation of motion \eqref{eq:eqmo}. Leaving only the terms linear in $\delta\varphi$ and separating the variables as $\delta\varphi(x,t) = \eta(x)\cos\left(\omega t\right)$, we obtain a spectral problem of the type of time-independent Schr\"odinger equation:
\begin{equation}\label{eq:Schrodinger}
	\left[-\frac{d^2}{dx^2} + U(x)\right]\eta(x) = \omega^2 \eta(x),
\end{equation}
where the potential $U(x) = \left.\displaystyle\frac{d^2V}{d\varphi^2}\right|_{\varphi_{\rm K}^{}(x)}$ is called ``stability potential'' or sometimes ``quantum-mechanical potential''. (Notice that here we do not consider any pathological cases of the potential $V(\varphi)$, considering it to be quite good.)

It can be easily shown that the discrete spectrum of the Schr\"odinger operator in \eqref{eq:Schrodinger} always has a zero eigenvalue corresponding to the zero mode (arising due to translational invariance) with the eigenfunction being
\begin{equation}\label{eq:phiprime}
    \eta_0^{}(x) = \varphi_{\rm K}^{\prime}(x).
\end{equation}
The square integrability of the function $\eta_0^{}(x)$ follows from the fact that the kink energy is finite. Thus, the eigenvalue $\omega_0^{}=0$ belongs to the discrete spectrum. Note that from \eqref{eq:Schrodinger} at $\omega_0^{}=0$, taking into account \eqref{eq:phiprime}, it immediately follows that the stability potential can be expressed in terms of the derivatives of the kink:
\begin{equation}\label{eq:quantum-mechanical_potential_kink_derivatives}
	U(x) = \frac{\varphi_{\rm K}^{\prime\prime\prime}(x)}{\varphi_{\rm K}^{\prime}(x)}.
\end{equation}

From the above it follows that the asymptotic behavior of the stability potential \eqref{eq:quantum-mechanical_potential_kink_derivatives}, as well as of the zero mode \eqref{eq:phiprime}, can be easily found for deformed kinks. For example, for the power-law asymptotics \eqref{eq:f-deformed_kink_asymptotics} we immediately obtain the power-law asymptotics of the zero mode and the stability potential as $x\to\pm\infty$: $\eta_0^{}(x)\sim 1/(2x^2)$ and $U(x)\sim 6/x^2$. The function $\eta_0^{}(x)$ is always nodeless, which is a consequence of the monotonicity of the kink, therefore, zero mode is the lowest level of the discrete spectrum. Taking into account that $U(x)\to 0$ as $x\to\infty$ and hence the zero mode lies on the boundary of the continuous spectrum, we conclude that the kink with asymptotics \eqref{eq:f-deformed_kink_asymptotics} does not have vibrational modes, see also ref.~\cite{Gani.JPCS.2020.no-go}.

In the case of super-exponential asymptotics \eqref{eq:f-1-deformed_kink_asymptotics} the stability potential behaves as $U(x)\sim e^{4x}$ at $x\to +\infty$. while for the super-super-exponential asymptotics \eqref{eq:f-1f-1-deformed_kink_asymptotics} we have the super-exponential explosive growth of the stability potential at $x\to +\infty$: $U(x) \sim e^{e^x}$. On the other hand, to the slow logarithmic asymptotics \eqref{eq:ff-deformed_kink_asymptotics} corresponds the following asymptotic behavior of the stability potential: $U(x) \sim 2/x^2$ with logarithmic corrections at $x\to +\infty$.

\section{Conclusion}
\label{sec:Conclusion}

In this letter, we have considered some aspects of applying the deformation procedure to field-theoretic models having kink solutions. Our goal was to understand how the asymptotic behavior of the potential and kink changes when using deforming functions of a certain type. Namely, we assumed that the deforming function is strictly monotonically increasing, vacua of the model are the fixed points of this function, and the function is singular --- in a certain way it goes to infinity in vacua. We also discussed the possibility of deformations using a function that is the reverse of the described one.

We have shown that with such deformations, the asymptotics of both the model potential and the kink significantly change. In particular, we have demonstrated how, starting with a model having kinks with exponential asymptotics, it is possible to obtain models having kinks with power-law, logarithmic, super-exponential and super-super-exponential asymptotics.  We have also explained how asymptotic behavior of the stability potential and the zero mode can be obtained for any deformed kink.

This study is part of an actively developing area of research, see the section ``Introduction'' for details. A number of interesting questions, however, remained outside the scope of this paper and could become the subject of separate publications.
\begin{itemize}
    \item It would be interesting to consider generalizations of the obtained results to more general cases of deformation function. For example as a first step, one could consider
    \begin{equation}
        f(\varphi)\approx\varphi_0^{} - C \exp\left[-\displaystyle\frac{\varepsilon}{(\varphi_0^{}-\varphi)^m}\right]
    \end{equation}
    with $C>0$, $\varepsilon>0$ and integer $m\ge 2$.
    \item As has already been mentioned above, successive application of the deformation function $f^{-1}$, accelerates the kink asymptotics, and it can be assumed that in the limit of infinitely large number of successive deformations we obtain a compact kink. In the future, it would be interesting to understand how to consider such a limit.
\end{itemize}

\section{Acknowledgments}

The work of VAG was partially supported within the state assignment of the Ministry of Science and Higher Education of the Russian Federation, project No.\ FSWU-2023-0031. AMM expresses gratitude to the Islamic Azad University Quchan branch for their generous grant. The authors are also grateful to the anonymous referee for his/her useful criticism, questions and suggestions, which helped to improve presentation of the study.

\end{document}